\documentclass[%
 reprint,
showkeys,
 amsmath,amssymb,
 aps,
longbibliography,
floatfix,
]{revtex4-1}

\usepackage{graphicx}
\usepackage{dcolumn}
\usepackage{bm}
\usepackage{hyperref}

\usepackage{color}
\usepackage[dvipsnames]{xcolor}
\usepackage{amsmath}
\usepackage{amssymb}
\usepackage{siunitx} 
\usepackage{wasysym}
\usepackage[titletoc,toc,title]{appendix}
\usepackage{setspace}
\usepackage{natbib}
\usepackage{subfigure}
\usepackage{float}
\usepackage{cancel}
\usepackage[normalem]{ulem}

\newcommand{\eps}{\varepsilon}
\newcommand{\ifl}{^{(i)}}
\newcommand{\efl}{^{(e)}}

\date{\today}

\begin{document}
\title{Controlling Dispersive Hydrodynamic Wavebreaking in a Viscous Fluid Conduit}

\author{Dalton V. Anderson}
\altaffiliation[Also at ]{Department of Aerospace Engineering, University of Colorado, Boulder, CO 80302, USA.}
\author{Michelle D. Maiden}\email{Michelle.Maiden@Colorado.edu}
\author{Mark A. Hoefer}
\homepage{https://www.colorado.edu/amath/research/dispersive-hydrodynamics-lab}
\affiliation{Department of Applied Mathematics, University of Colorado, Boulder, Colorado 80302, USA}

\begin{abstract}
  The driven, cylindrical, free interface between two miscible, Stokes
  fluids with high viscosity contrast have been shown to exhibit
  dispersive hydrodynamics.  A hallmark feature of dispersive
  hydrodynamic media is the dispersive resolution of wavebreaking that
  results in a dispersive shock wave.  In the context of the viscous
  fluid conduit system, the present work introduces a simple,
  practical method to precisely control the location, time, and
  spatial profile of wavebreaking in dispersive hydrodynamic systems
  with only boundary control.  The method is based on tracking the
  dispersionless characteristics backward from the desired
  wavebreaking profile to the boundary.  In addition to the generation
  of approximately step-like Riemann and box problems, the method is
  generalized to other, approximately piecewise-linear dispersive
  hydrodynamic profiles including the triangle wave and N-wave.  A
  definition of dispersive hydrodynamic wavebreaking is used to obtain
  quantitative agreement between the predicted location and time of
  wavebreaking, viscous fluid conduit experiment, and direct numerical
  simulations for a range of flow conditions.  Observed space-time
  characteristics also agree with triangle and N-wave predictions.
  The characteristic boundary control method introduced here enables
  the experimental investigation of a variety of wavebreaking profiles
  and is expected to be useful in other dispersive hydrodynamic media.
  As an application of this approach, soliton fission from a large,
  box-like disturbance is observed both experimentally and
  numerically, motivating future analytical treatment.
\end{abstract}


\maketitle

\section{Introduction}\label{sec:Intro}
The interfacial dynamics of a viscous fluid conduit exhibit a wide
range of dispersive hydrodynamic behavior observable in other
geophysical, superfluidic, optical, and condensed matter media
\cite{el_dispersive_2016}.  A viscous fluid rises buoyantly through a
more viscous, stationary fluid and, under appropriate conditions, the
interface resembles that of a deformable pipe.  The interface's
dynamics can be accurately described by the conduit equation
\cite{olson_solitary_1986,lowman_dispersive_2013,whitehead_wave_1988},
a nonlinear dispersive partial differential equation that has been
shown to admit a rich zoology of multiscale coherent wave solutions
\cite{maiden_modulations_2016}.  Experimentally, solitons
\cite{scott_observations_1986,olson_solitary_1986,helfrich_solitary_1990},
dispersive shock waves (DSWs), and the interactions between them
\cite{lowman_interactions_2013,maiden_observation_2016,maiden_solitonic_2018}
have been observed to be in excellent agreement with conduit equation
predictions.  In this sense, the viscous fluid conduit is an ideal
laboratory environment for the study of dispersive hydrodynamics.

Dispersive shock waves are coherent structures that are fundamental to
dispersive hydrodynamics \cite{el_dispersive_2016}.  A DSW is
an expanding, oscillating train of amplitude-ordered nonlinear waves
composed of a large amplitude solitary wave adjacent to a
monotonically decreasing wave envelope that terminates with a packet
of small amplitude dispersive waves. 
DSWs result when nonlinear self-steepening is compensated by conservative wave dispersion in contrast to the viscous shock waves of dissipative media. 
The model problem for shock generation is the Riemann problem that consists of an initial, sharp step in amplitude.

A major challenge for the experimental study of DSWs is the controlled
realization of wavebreaking that involves the spontaneous generation
of oscillations when nonlinear self-steepening enhances small-scale
dispersive processes.  One obstacle to the laboratory generation of a
desired wavebreaking profile is its reliable initiation from only
boundary control.  Laboratory nonlinear dispersive wave environments
constrained by boundary control include fluids such as shallow water
wave tanks (see, e.g. \cite{trillo_observation_2016}) and viscous
fluid conduits \cite{scott_observations_1986}.  Also included are
non-fluid systems such as intense laser light propagation in optical
fibers \cite{xu_dispersive_2017}, magnetic spin waves
\cite{janantha_observation_2017}, and granular crystals
\cite{chong_nonlinear_2017}.  It can be difficult to achieve
controlled wavebreaking conditions without boundary interactions.
Here, we report on a simple mathematical observation that yields a
feasible way to achieve a variety of wavebreaking profiles away from
boundaries.  We track the (long-wave) characteristics of the
dispersionless conduit equation backwards in time from the desired
wavebreaking profile.  The resulting solution is then used as a
boundary condition for conduit experiments to realize a variety of
wavebreaking profiles at desired spatial locations.  This technique
was used to generate dispersive hydrodynamic flows in a viscous fluid
conduit \cite{maiden_observation_2016,maiden_solitonic_2018}.  In a
related work in laser propagation through a nonlinear fiber
\cite{wetzel_experimental_2016}, experimentalists created hyperbolic
simple waves aided by the corresponding dispersionless long-wave
shallow water model.

\begin{figure*}
  \centering
  \subfigure[~step]{\includegraphics{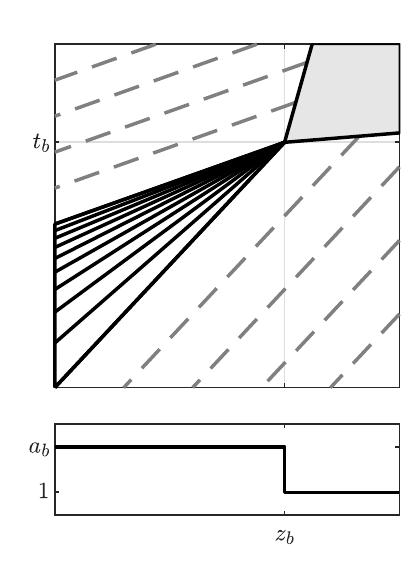}}
  \subfigure[~box]{\includegraphics{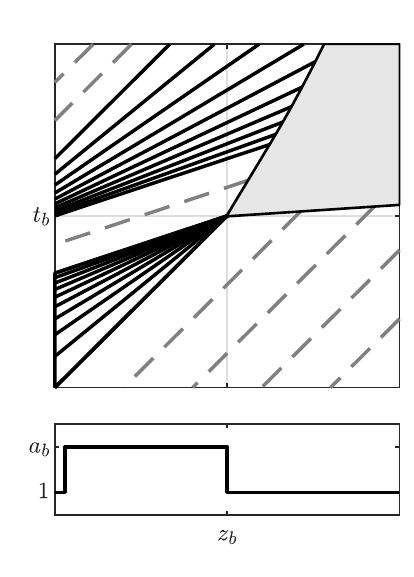}}
  \subfigure[~triangle]{\includegraphics{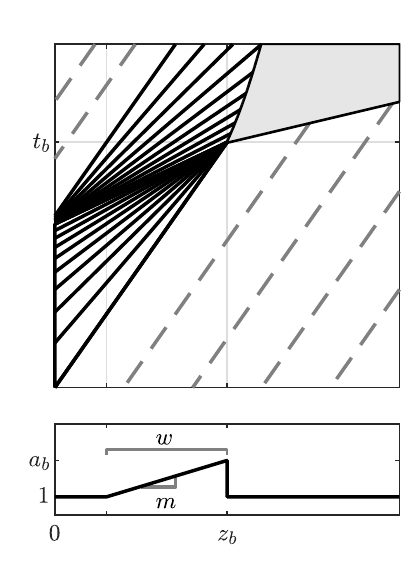}}
  \subfigure[~N-wave]{\includegraphics{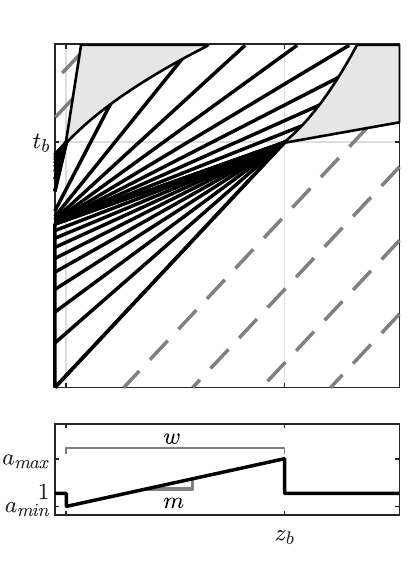}}
  \caption{Characteristic plots (large plots) and wavebreaking
    profiles at time of breaking (small plots). The gray regions are
    areas where wavebreaking has occured and small-scale dispersion is
    important so the inviscid Burger's solution is no longer
    valid. The light vertical and horizontal lines depict strong and
    weak discontinuities in the profiles at the time of
    breaking. Shown here are step (a), box (b), triangle (c), and
    N-wave (d) wavebreaking profiles.}
      \label{fig:chars2}
\end{figure*}
Changing between boundary and initial conditions is a useful tool in
the study of nonlinear waves.  The piston shock problem is a canonical
boundary value problem in the theory of classical shock waves.
G. B. Whitham reformulated the problem by tracing characteristics back
from the piston to an equivalent initial value problem that was then
solved via the method of characteristics
\cite{whitham_non-linear_1965}.  Here, we do the reverse by converting
a desired initial value problem into a boundary value problem in the
context of dispersive shock waves.  We use this approach to precisely
realize several wavebreaking profiles in experiment: step, box,
triangle, and N-wave configurations.  We find that, despite neglecting
short-wave dispersion, we can precisely control the location and time
of wavebreaking as well as the long-wave characteristics that lead up
to breaking.  This is supported both theoretically by numerical
simulations of the conduit equation and experimentally with our
viscous conduit setup.  This control method is also used to generate a
large number of solitary waves (soliton fission) from a box
profile. This relatively simple approach turns out to be quite
effective and could be applied to other dispersive hydrodynamic media.
Indeed, dispersive hydrodynamic wavebreaking and its control have
played a decisive role in recent shallow water experiments
\cite{trillo_experimental_2016,trillo_observation_2016} and intense
light propagation in defocusing, optical media
\cite{ghofraniha_measurement_2012,wetzel_experimental_2016}.

The paper is organized as follows. Section \ref{sec:Theory} is the theoretical section and includes background information on the conduit equation and the mathematical procedure for converting the initial value problem into a boundary value problem. Section \ref{sec:Exp} is the experimental section and covers both numerical and experimental methods and analysis. Conclusions are in Section \ref{sec:Con}.

\section{Theory}\label{sec:Theory}
\subsection{Conduit Equation}
Conduits generated by the low Reynolds number, buoyant dynamics of two miscible fluids with differing densities and viscosities were first studied in the context of geological and geophysical processes \cite{whitehead_dynamics_1975}. 
We create viscous fluid conduits with glycerine as the exterior fluid and dyed, diluted glycerine as the interior fluid. 
Long-wave, slowly varying perturbations to a uniform background conduit result in the conduit equation \cite{olson_solitary_1986,lowman_dispersive_2013}
\begin{equation}\label{eq:conduit}
  A_T + \frac{g\Delta}{8\pi\mu\ifl}\left(A^2\right)_Z - \frac{\mu\efl}{8\pi\mu\ifl}\left(A^2\left(A^{-1}A_T\right)_Z\right)_Z = 0,
\end{equation}
\noindent 
where $\mu\ifl$ is the interior dynamic viscosity, 
$\mu\efl$ is the exterior dynamic viscosity, 
$\Delta=\rho\efl-\rho\ifl$ is the difference in exterior to interior fluid densities, and $g$ is gravitational acceleration.
This equation approximately governs the evolution of the circular interface with cross-sectional area $A$ at vertical height $Z$ and time $T$. 
The simplest wavebreaking configuration is a step decrease in conduit area 
\begin{equation}\label{eq:GP}
    A(Z,T_b) = \begin{cases}
                A_2, & Z < Z_b \\
                A_1, & Z\geq Z_b 
             \end{cases},
\end{equation}
for some $A_2>A_1=\pi R_0^2$, where $R_0$ is the conduit radius, $Z_b$
is the breaking location, and $T_b$ is the breaking time.  We can
nondimensionalize the equation and rescale the leading area to unity
via the scalings
\begin{equation}\label{eq:scaling1}
  \begin{array}{ccc}
    a=\frac{1}{\pi R_0^2}A, & z=\frac{\sqrt{8\eps}}{R_0}Z,  & t = \frac{gR_0\Delta\sqrt{\eps}}{\sqrt{8}\mu\ifl}T,
  \end{array}
\end{equation}
where $\eps=\frac{\mu\ifl}{\mu\efl}$ is the interior to exterior
viscosity ratio.  Then, the conduit equation in nondimensional form is
\begin{align}\label{eq:conduitND}
    a_t + (a^2)_z -(a^2(a^{-1}a_t)_z)_z = 0, \, z\in\mathbb{R}, \, t>t_b.
\end{align}
We represent the desired wavebreaking profile via the data $a(z,t_b) =
a_0(z)$. For example the step profile \eqref{eq:GP} is
\begin{align}\label{eq:GPscaled}
  a_0(z) = \begin{cases}
    a_b, & z<z_b \\
    1, & z\geq z_b
             \end{cases}, z\in\mathbb{R}.
\end{align}
where $a_b = A_2/A_1 > 1$ is the jump ratio and $z_b$, $t_b$ are the nondimensional breaking height and time, respectively. 
We also consider the box profile of amplitude $a_b$ and width $w$,
\begin{align}\label{eq:boxscaled}
    a_0(z) = \begin{cases}
                a_b, & z_b-w<z<z_b \\
                1,   & \textrm{else}
             \end{cases}, z\in\mathbb{R},
\end{align}
the triangle profile of amplitude $a_b$, width $w$, and hypotenuse slope $m=\frac{a_b-1}{w}$,
\begin{align}\label{eq:triscaled}
     a_0(z) = \begin{cases}
                mz + (a_b-mz_b), & z_b-w<z<z_b \\
                1,   &  \textrm{else}
             \end{cases}, z\in\mathbb{R},
\end{align}
and the N-wave profile of maximum amplitude $a_{max}$, minimum amplitude $a_{min}$, width $w$, and slope $m=\frac{a_{max}-a_{min}}{w}$,
\begin{align}\label{eq:Nscaled}
    a_0(z) = \begin{cases}
                mz + (a_b-mz_b), & z_b-w<z<z_b \\
                1,   &  \textrm{else}
             \end{cases}, z\in\mathbb{R}.
\end{align}
For example profiles, see Fig. \ref{fig:chars2}.
\subsection{Inviscid Burgers Equation}
\label{sec:Inv-Bur-EQ}

In order to approximately realize the breaking confirgurations that
give rise to $a_0(z)$, e.g. \eqref{eq:GPscaled}--\eqref{eq:Nscaled},
we seek to identify the spatio-temporal profile for times
\textit{prior} to breaking $t<t_b$.  We assume that prior to DSW
formation, the third order dispersive term is negligible.  A
dimensional analysis of eq.~(\ref{eq:conduit}) shows that the
nonlinear advective term dominates the dispersive term prior to
breaking if
\begin{equation}
  \label{eq:2}
  Z_b \gg \frac{\mu\efl}{g \Delta T_b},
\end{equation}
where $Z_b$ and $T_b$ are the dimensional breaking height and time,
respectively.  For the experiments reported here, $\mu\efl/( g
\Delta)\approx \SI{0.30}{\second\centi\meter}$, $T_b \in (80,140) \,
\SI{}{\second}$, and $Z_b \in (15,27)\, \SI{}{\centi\meter}$.
Therefore, we are well within the expected regime of validity.  We
will further justify the assumption of dispersionless dynamics with
numerical and physical experiments.  This is a long-wave assumption
that is valid when $|a_z|/|a|\sim |a_t|/|a|\ll 1 $, when nonlinearity
exceeds wave dispersion.  We therefore neglect the dispersive term,
$(a^2(a^{-1}a_t)_z)_z$ in \eqref{eq:conduitND}, and reverse time and
shift space via
    \begin{equation} \label{eq:coordSub0}
        \begin{array}{ccc}
            z = \zeta+z_b, & t = -(\tau-t_b), & a=u,
        \end{array}
    \end{equation}
where $u(\zeta,\tau)$ now satisfies the time-reversed inviscid Burgers Equation,
    \begin{align}\label{eq:IBEreverse}
        u_\tau - \left(u^2\right)_\zeta = 0, \tau>0, \zeta\in\mathbb{R} \\
        u(\zeta,0) = u_0(\zeta) = a_0(\zeta - z_b),
    \end{align}
which has the implicit solution
    \begin{equation}\label{eq:IBEsoln}
        u(\zeta,\tau) = u_0(\zeta + 2 u(\zeta,\tau)\tau).
    \end{equation}
Then, converting \eqref{eq:IBEsoln} back to conduit equation variables and evaluating at the boundary $z=0$, we have an implicit form of the boundary condition in terms of the known initial condition $a_0(z)$,
    \begin{equation}\label{eq:IBEsolnA}
        a(0,t) = a_0(-2a(0,t)\,t).
    \end{equation}
A necessary condition that precludes breaking for $0<t<t_b$ is $a_0'(z)<1/2t_b$ for all $z$.
    
\begin{figure}
    \centering
       \subfigure[~Boundary temporal profile.]{\includegraphics{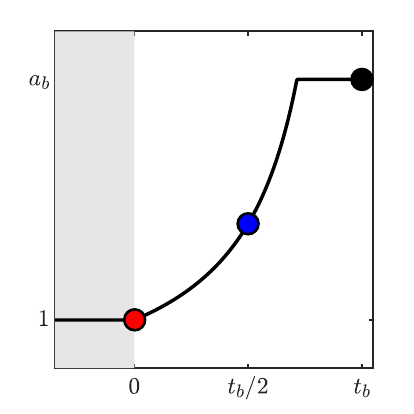}}
	   \subfigure[~Spatio-temporal development of the wavebreaking profile.]{\includegraphics{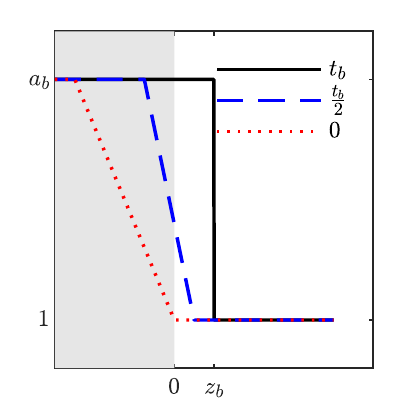}}     
    \caption{(a) Temporal profile of the boundary condition Eq.~\eqref{eq:BCDSW}. (b) Evolution of the rescaled rarefaction wave Eq.~\eqref{eq:BCDSW}. As time moves forward, the wave approaches the desired step. The dots in (a) correspond to the times depicted in (b), from left to right. }
    \label{fig:rare_wave_inv}
\end{figure}
\begin{figure}
  \centering
  \includegraphics{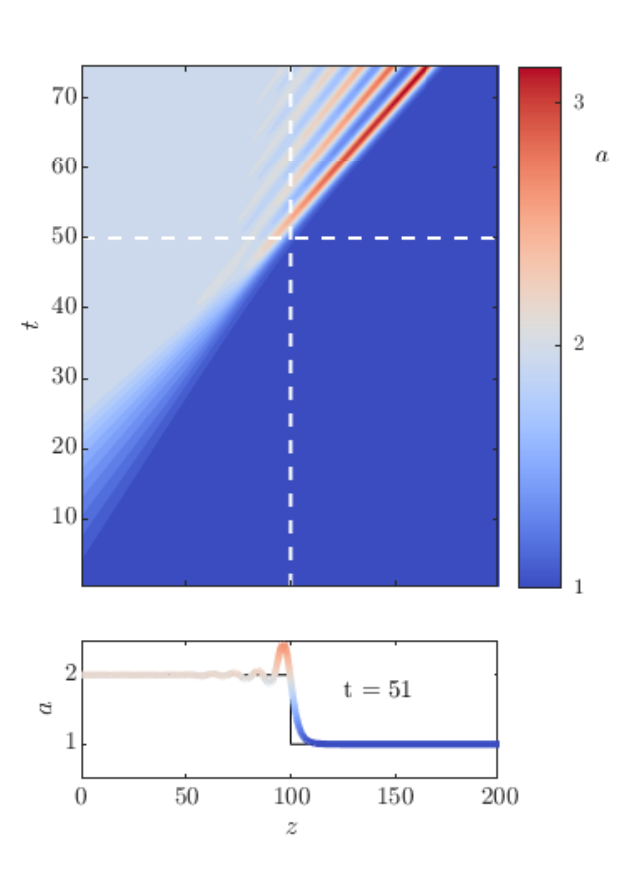}
  \caption{Numerical simulation of the conduit equation with initial
    condition $a(z,0)=1$ and the boundary condition equation
    \eqref{eq:BCDSW} with $z_b=100$ and $a_b=2$. The predicted $z_b =
    100$ and $t_b=50$ are identified by the intersection of the
    vertical and horizontal dashed lines.  The oscillatory wave
    profile at the extracted time and location of breaking $t_b = 51$,
    $z_b = 97$ (see Sec.~IIC) and its corresponding theoretical step
    profile are shown in the lower panel.}
  \label{fig:numerics_contour}
\end{figure}
We now consider the specific case of step data \eqref{eq:GPscaled}.
For this case, the self-similar, rarefaction wave solution is operable
\begin{equation}\label{eq:raresoln}
  u(\zeta,\tau) = \begin{cases}
    u_- & : \zeta\le 2u_-\tau,  \\
    \frac{\zeta}{2\tau} & : 2u_-\tau \le \zeta \le 2u_+\tau, \\
    u_+ & :\zeta\ge 2u_+\tau .
  \end{cases}
\end{equation}  
The substitution \eqref{eq:coordSub0} along with
\begin{equation} \label{eq:coordSub1}
  \begin{array}{ccc}
    t_b = \frac{z_b}{2}\frac{a_b-1}{a_b},  & u_- = 1,  & u_+ = a_b 
  \end{array}
\end{equation}
yields the sought for boundary condition
\begin{equation}\label{eq:BCDSW}
  a(0,t) = 
  \begin{cases}
    1               & :      t     \le 0                         \\
    (1-2t/z_b)^{-1} & :  0 < t <   \frac{(a_b - 1)}{2a_b} z_b    \\
    a_b & : t \ge \frac{(a_b - 1)}{2a_b} z_b
  \end{cases}.
\end{equation}
This solution and its evolution are shown in
Fig. \ref{fig:rare_wave_inv}. Note that we have chosen the specific
breaking time $t_b$ in \eqref{eq:coordSub1} so that $a(0,t)=1$ for
$t\leq 0$. Any desired breaking time can be achieved by a simple time
shift.  

\subsection{Definition of Dispersive Wavebreaking Point}
\label{sec:defin-disp-wavebr}

In order to compare the predicted versus actual dispersive
wavebreaking profile, we have performed direct numerical simulations
of the conduit equation (\ref{eq:conduitND}) with the boundary
condition (\ref{eq:BCDSW}) for $z_b = 100$, $t_b = 50$, and $a_b = 2$.
A space-time contour plot of the simulation and its comparison with
the predicted step profile are shown in
Fig.~\ref{fig:numerics_contour}.  Near the point of breaking,
dispersion is no longer negligible; as a result, a perfect Riemann
step is not realized in the conduit equation. Therefore, we introduce
a definition of dispersive wavebreaking as follows.

\begin{figure}
  \centering
  \includegraphics{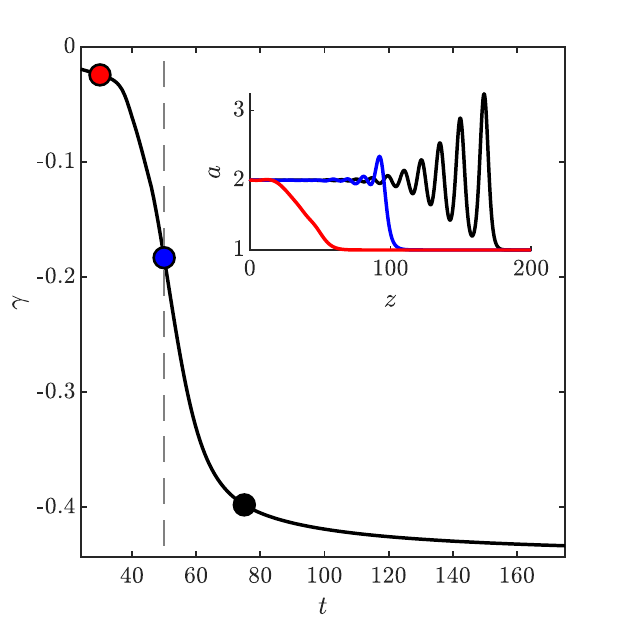}
  \caption{Leading edge slope $\gamma(t)$ in eq.~(\ref{eq:3})
    extracted from the numerical simulation depicted in
    Fig.~\ref{fig:numerics_contour} and the inset.  The time of
    breaking $t_b$ occurs at the inflection point $\ddot{\gamma}(t_b)
    = 0$ of the slope as a function of time (middle circle).  Inset:
    three profiles in space corresponding to the marked points in
    time. The predicted breaking time is $t_b=50$.}
  \label{fig:numerics_breaking}
\end{figure}
By an analysis of numerical simulations of the conduit equation
\eqref{eq:conduitND} for a variety of breaking points $(z_b,t_b)$ and
amplitudes $a_b$, we introduce a robust definition of the space-time
location of wavebreaking from numerical simulations and experiment
using the slope, $\gamma$, at the wave front's midpoint
\begin{equation}
  \label{eq:3}
  \begin{split}
    &\gamma(t) = \left . \frac{\partial a}{\partial z}(z,t) \right
    |_{z = z_m(t)} , \quad \mathrm{where} ~ z_m(t) ~
    \mathrm{satisfies} \\
    &a(z_m,t) = \frac{1}{2} \left ( \max_z a(z,t) \right ) +
    \frac{1}{2} \left ( \min_z a(z,t) \right ) .
  \end{split}
\end{equation}
We define the breaking time for both numerical simulation and
experiment as the time $t_b$ when the slope achieves an inflection
point in time: $\ddot{\gamma}(t_b) = 0$.  We then define the breaking
height $z_b$ as the location where the profile achieves its maximum
amplitude: $a(z_b,t_b) = \max_z a(z,t_b)$.  The breaking point
$(z_b,t_b)$ identified by the dashed lines in
Fig.~\ref{fig:numerics_contour} corresponds to the inflection point of
the slope $\gamma(t)$ evolution shown in
Fig.~\ref{fig:numerics_breaking}.

\subsection{Generalizations to Piecewise Linear Profiles}\label{sec:Gen-Tech}
    This method of neglecting the dispersive term can be used to generate a variety of initial conditions, the formulae for which are included in the appendix\ref{appx:BCs}. In Fig.~\ref{fig:chars2}, we show characteristic plots based on the dispersionless approach to generate a step in Fig.~\ref{fig:chars2}(a), a box in Fig.~\ref{fig:chars2}(b), a triangle in Fig.~\ref{fig:chars2}(c), and an N-wave in Fig.~\ref{fig:chars2}(d). 
    
\begin{figure}
  \centering
  \subfigure[]{\includegraphics[width=1.65in]{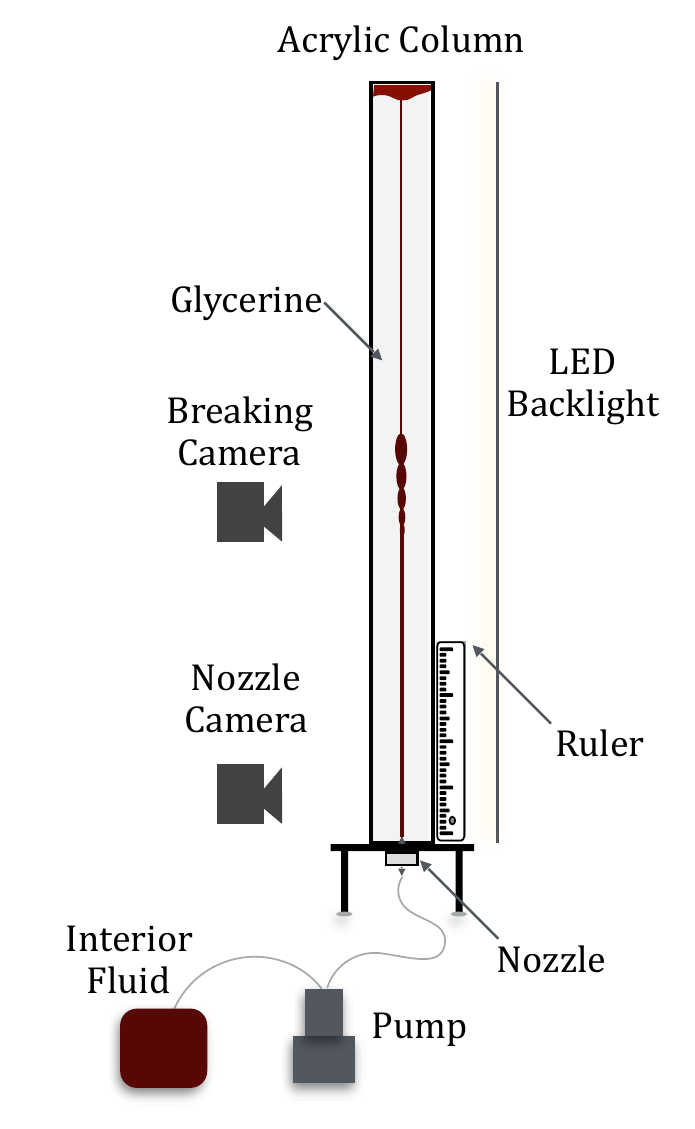}}
  \subfigure[]{\includegraphics[width=1.65in]{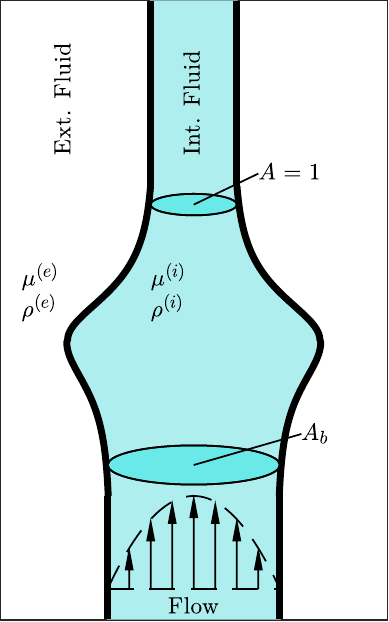}}
  \caption{(a) Schematic of the experimental setup. (b) Schematic of
    the conduit near $t=t_b$. Dispersion leads to the emergence of a
    nonmonotonic profile near the time of breaking.}
  \label{fig:expsys}
\end{figure}
    There are some restrictions on the types of profiles we can generate. In order for an entire profile to be above the $z=0$ boundary at the time of breaking, we require the width of the profile to be less than or equal to $z_b$. For triangle waves and N-waves, there is an additional width-height ratio that must be satisfied in order for the diagonal portions to be fully realized. These conditions are listed in the appendix\ref{appx:BCs}.
\section{Experiment}\label{sec:Exp}
\subsection{Setup}
The experimental apparatus shown in Fig. \ref{fig:expsys}(a) consists
of a square acrylic column with dimensions $\SI{4}{\centi\meter}
\times \SI{4}{\centi\meter} \times \SI{92}{\centi\meter}$; the column
is filled with glycerine a highly viscous, transparent, exterior
fluid.  A nozzle is installed at the base of the column to allow for
the injection of the interior fluid.  To eliminate surface tension
effects, the interior fluid is a solution of glycerine, deionized
water, and black food coloring.  As a result, the interior fluid has
both lower viscosity and density than the exterior fluid $(\mu^{(i)}
\ll \mu^{(e)}$ , $\rho^{(i)} < \rho^{(e)})$ and we assume mass
diffusion is negligible.

Interior fluid is drawn from a separate reservoir and injected through
the nozzle via a high precision computer controlled piston pump.  The
interior fluid rises buoyantly.  By injecting at a constant rate, a
vertically uniform fluid conduit is established.  This uniform steady
state is referred to as the background conduit, and is
well-approximated as pipe (Poiseuille) flow, verified in
\cite{maiden_observation_2016}.  Data acquisition is performed using
high resolution cameras equipped with macro lenses at the injection
nozzle and the predicted breaking height.  A ruler is positioned
beside the column within camera view for calibration purposes and to
determine the observed breaking height.

\begin{figure}
  \centering
  \subfigure[]{\includegraphics{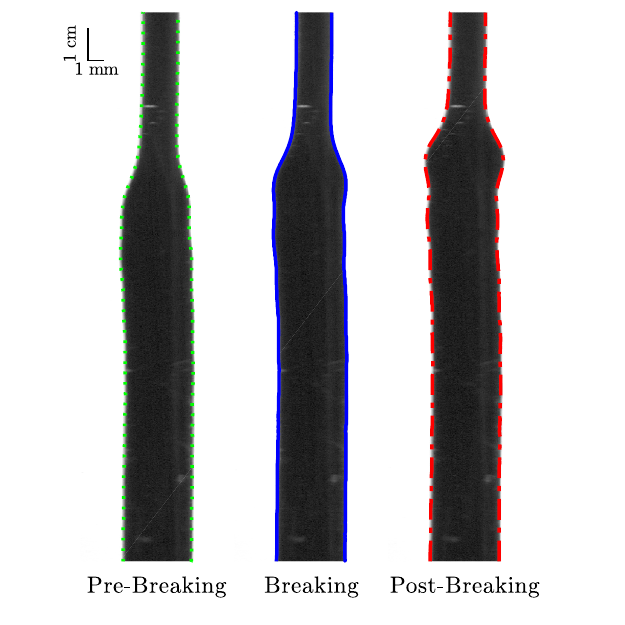}}
  \subfigure[]{\includegraphics{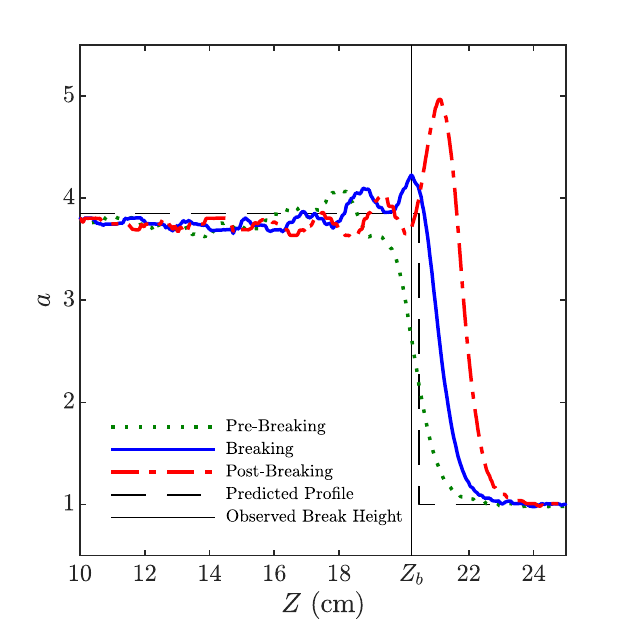}}
  \caption{(a) Processed images from a glycerine trial. Measured
    parameter values are $\mu\ifl=72\pm1 \si{cP}$,
    $\rho\ifl=1.222\pm0.001 \si{g/cm^3}$, $\mu\efl=1190\pm20 \si{cP}$,
    $\rho\efl=1.262\pm0.001 \si{g/cm^3}$, and
    $Q_0=0.25\pm0.01 \si{ml/min}$.  The grayscale images are overlayed
    with the extracted conduit edges.  (b) Nondimensional area plot
    corresponding to the images in (a). The vertical line indicates
    the desired step front that results from the procedure in Section
    \ref{sec:Methods}. The dashed line indicates the expected step in
    a dispersionless system. Predictions were fit to the found
    Poiseuille flow relation \eqref{eq:Poiseuille}.}
  \label{fig:BreakEdge}
\end{figure}

\begin{figure*}
  \centering
  \subfigure[]{ \includegraphics{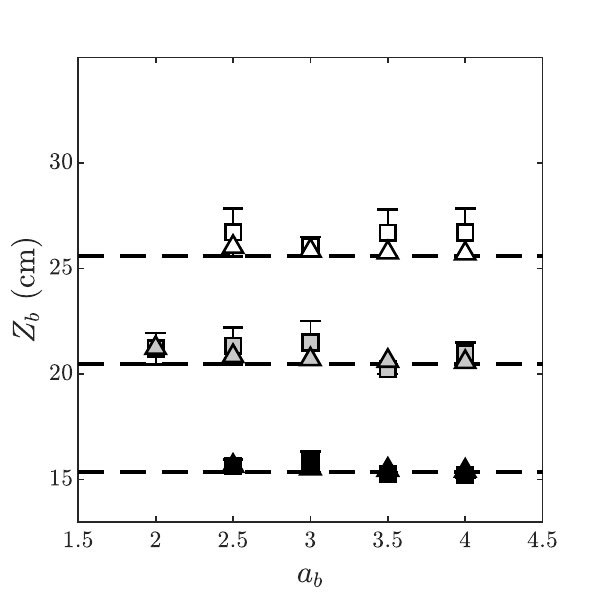}}
  \subfigure[]{ \includegraphics{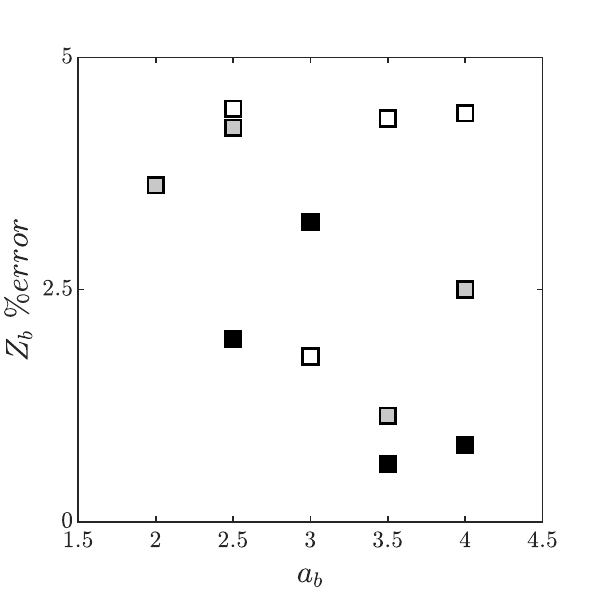}}
  \subfigure[]{ \includegraphics{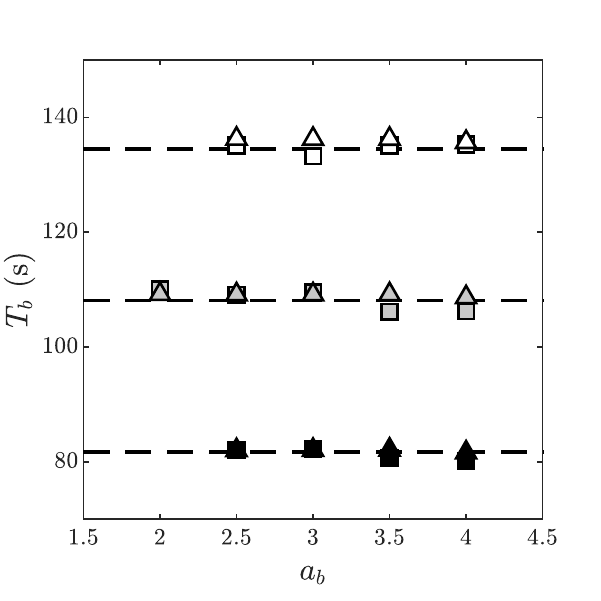}}
  \subfigure[]{ \includegraphics{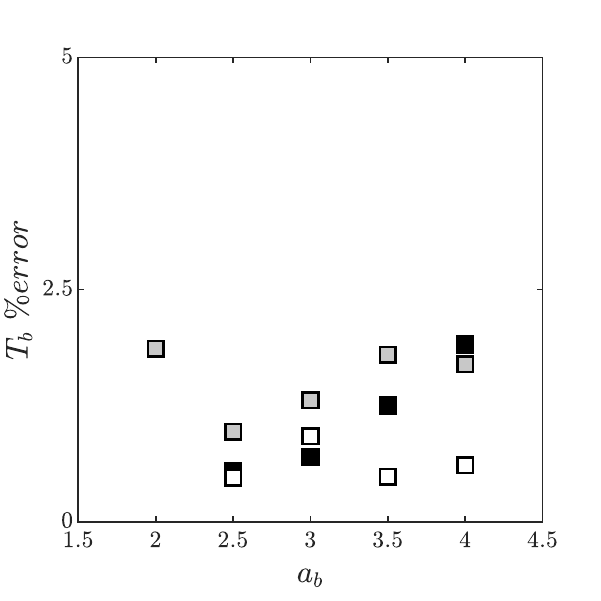}}
  \caption{Comparison of dispersionless (long-wave) theory (dashed
    lines), full conduit equation numerics (triangles) and glycerine
    experiments (squares) for the step wavebreaking configuration. (a)
    Breaking height results and (b) relative error for experiments as
    a function of jump ratio $a_b$, with the same fluid parameters as
    those in Fig.~\ref{fig:BreakEdge}. (c) Breaking time results and
    (d) relative error for those same experiments. Note the breaking
    time error bars are smaller than the symbols used. The black
    squares correspond to an expected $z_b = \SI{15.3}{\centi\meter}$,
    the gray to $z_b = \SI{20.5}{\centi\meter}$, and the white to $z_b
    = \SI{25.6}{\centi\meter}$.}
  \label{fig:BreakHeightGlycerin}
\end{figure*}

\subsection{Methods}\label{sec:Methods}
In order to use the results from Section \ref{sec:Theory}, we rescale
\eqref{eq:BCDSW} from the nondimensional conduit equation (lower case
variables) to physical parameters (upper case variables).  Following
the scaling in \eqref{eq:scaling1} and the Poiseuille flow relation,
\begin{equation}\label{eq:Poiseuille}
  Q = \frac{\pi g \Delta}{ 8\mu\ifl } R^{4},
\end{equation}
where $R$ is the conduit radius, a volumetric flow rate profile $Q(t)$
is generated for the desired wavebreaking configuration; see the
appendix\ref{appx:BCs}.  The camera near the nozzle takes images
before and after the initiation of the boundary volumetric flow
profile, so background conduit diameters are measured.  The breaking
camera takes several high-resolution images before, during, and after
the time of breaking.  A schematic of the conduit near the height and
time of breaking is shown in Fig. \ref{fig:expsys}(b).  After breaking
occurs, the pump is reduced to the background rate $Q_0$, and the
conduit is left to equilibrate before the next trial while fluid is
extracted from the top of the fluid column.
	
\begin{figure}[h]
  \centering
  \subfigure[]{\includegraphics{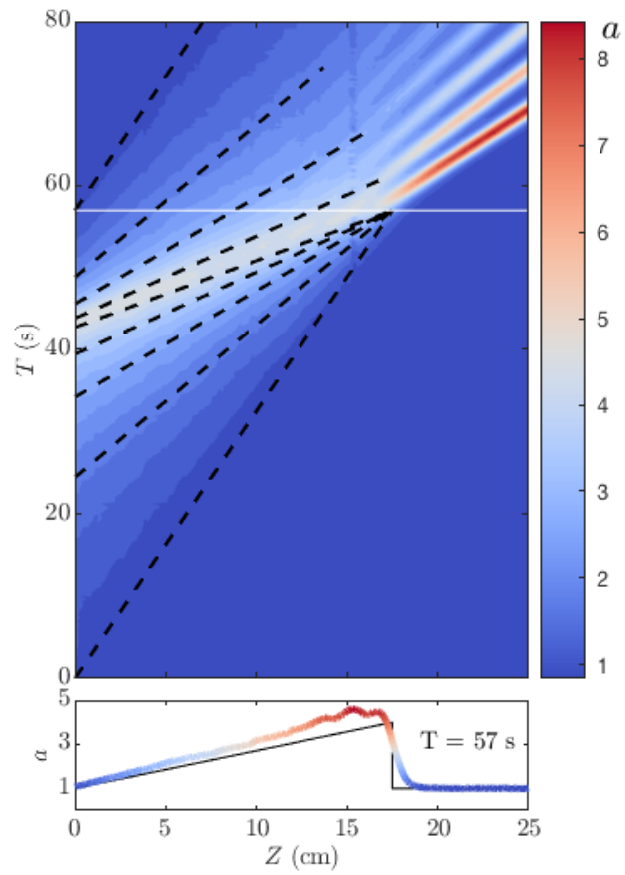}}
  \subfigure[]{\includegraphics{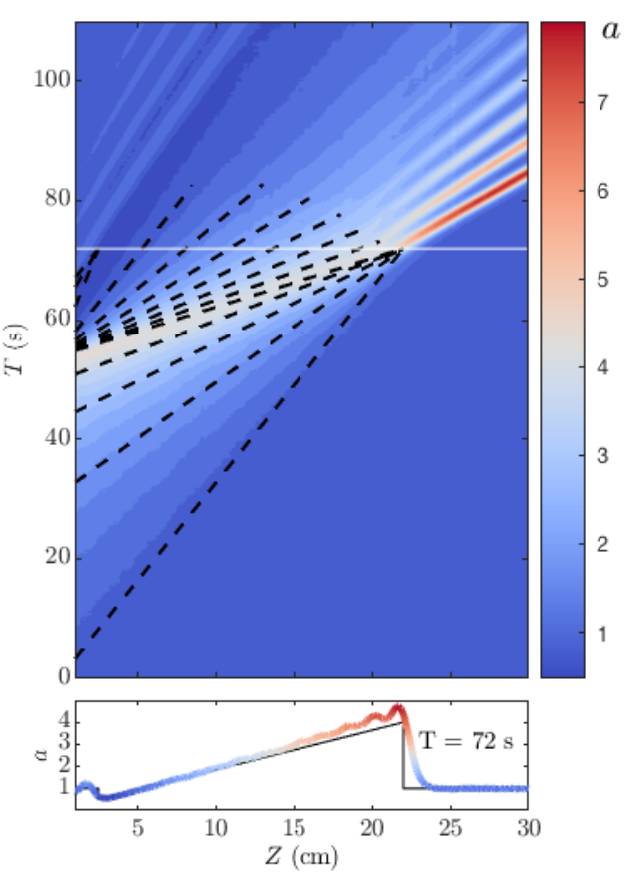}}
  \caption{Experimental data for triangle wave (top) and N-wave
    (bottom) boundary conditions. Overlay lines: fitted characteristic
    data for triangle (a) and N-wave (b) profiles.}
  \label{fig:GeneralizedCharExpt}
\end{figure}

The images from the camera are processed in \textsc{Matlab} to extract
the conduit edges as shown in Fig.~\ref{fig:BreakEdge}(a) by taking a
horizontal row of pixels and calculating the maximum and minimum
derivatives for each row.  As the background is white and the conduit
is black, this identifies the conduit boundary.  The edge data is then
processed with a low-pass filter to reduce noise due to pixelation of
the photograph and any impurities (such as bubbles) in the exterior
fluid.  The conduit diameter is then calculated as the number of
pixels between the two edges.
    
We calibrate the ratio $\mu\ifl/\Delta$ by fitting the observed
diameter data and its corresponding nominal volumetric flow rate to
the Poiseuille flow relation \eqref{eq:Poiseuille}.  We then determine
the experimental breaking height $z_b$ and time $t_b$ using the
definition in Section \ref{sec:defin-disp-wavebr}.  Experimental
profiles of pre-breaking, breaking, and post-breaking compared with
the desired step profile are shown in Fig.~\ref{fig:BreakEdge}(b).

\subsection{Results for Step Profile}

Utilizing the theoretically prescribed boundary condition \eqref{eq:1}
that results in the step profile \eqref{eq:GPscaled} at $(z,t) =
(z_b,t_b)$ for the solution of the dispersionless equation $a_t +
2aa_z = 0$ with unit step initial condition, we perform numerical
simulations of the full conduit equation \eqref{eq:conduitND} and
carry out fluid experiments.  A typical numerical simulation is
depicted in the contour plot of Fig.~\ref{fig:numerics_contour}.  The
wave profile steepens until dispersion becomes important, coinciding
with the emergence of oscillations, which prevent the formation of a
discontinuity.  Nevertheless, using the method described in the
previous section to extract $(z_b,t_b)$ from the simulations, the
observed breaking locations and times are within 3.75\% and 1.35\%
relative error, respectively of their predicted values across a range
of step ratios and breaking heights as depicted by the triangles and
dashed lines in Fig.~\ref{fig:BreakHeightGlycerin}.

For the experiment, thirteen trials were taken over the course of four
hours. The main results of this experiment are also shown in
Fig.~\ref{fig:BreakHeightGlycerin}.  The predicted breaking heights
and times ($z_{b,in}$ and $t_{b,in}$) (denoted with dashed lines) are
very close to the experimentally observed values ($z_{b,out}$ and
$t_{b,out}$) (denoted by squares).  Figure
\ref{fig:BreakHeightGlycerin} includes the theoretical prediction,
numerical simulations, and experiment for the breaking height (a) and
the breaking time (b).  All experiments were under $5\%$ relative
error in breaking height $z_b$ and $2.5\%$ relative error in breaking
time $t_b$.  Therefore, a high degree of wavebreaking control is
achieved by our approach.  Furthermore, the prediction's accuracy
appears to be independent of the breaking amplitude $A_b$, which is
consistent with the dimensional analysis requirement (\ref{eq:2}) that
is independent of wave amplitude.

These observations extend previous experimental comparisons of
theoretical predictions for the conduit equation involving solitons
\cite{scott_observations_1986,olson_solitary_1986,helfrich_solitary_1990,lowman_interactions_2013}
and dispersive shock waves
\cite{maiden_observation_2016,maiden_solitonic_2018}---i.e.,
nonlinear, \textit{dispersive} waves---into the non-dispersive,
nonlinear regime.  It is noteworthy that theoretical predictions
derived from the relatively simple inviscid Burgers model,
$a_t + 2aa_z = 0$, agree so well with numerical simulations and
experiment across a range of parameter values.
  
\begin{figure}
  \centering
  \subfigure[]{\includegraphics{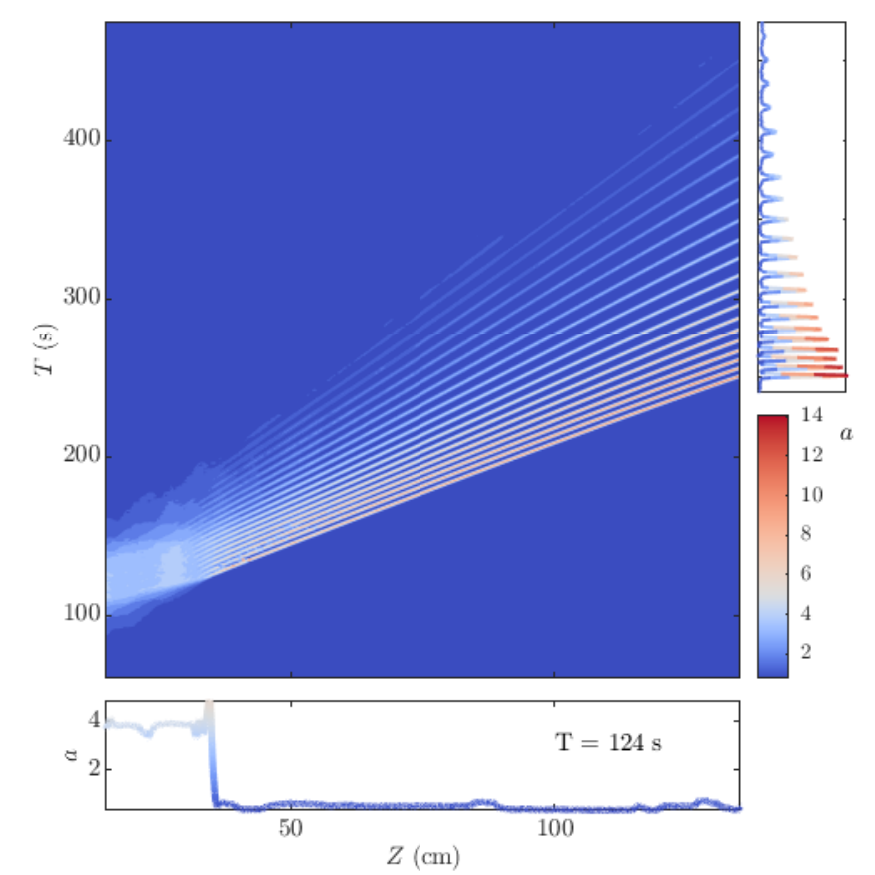}}
  \subfigure[]{\includegraphics{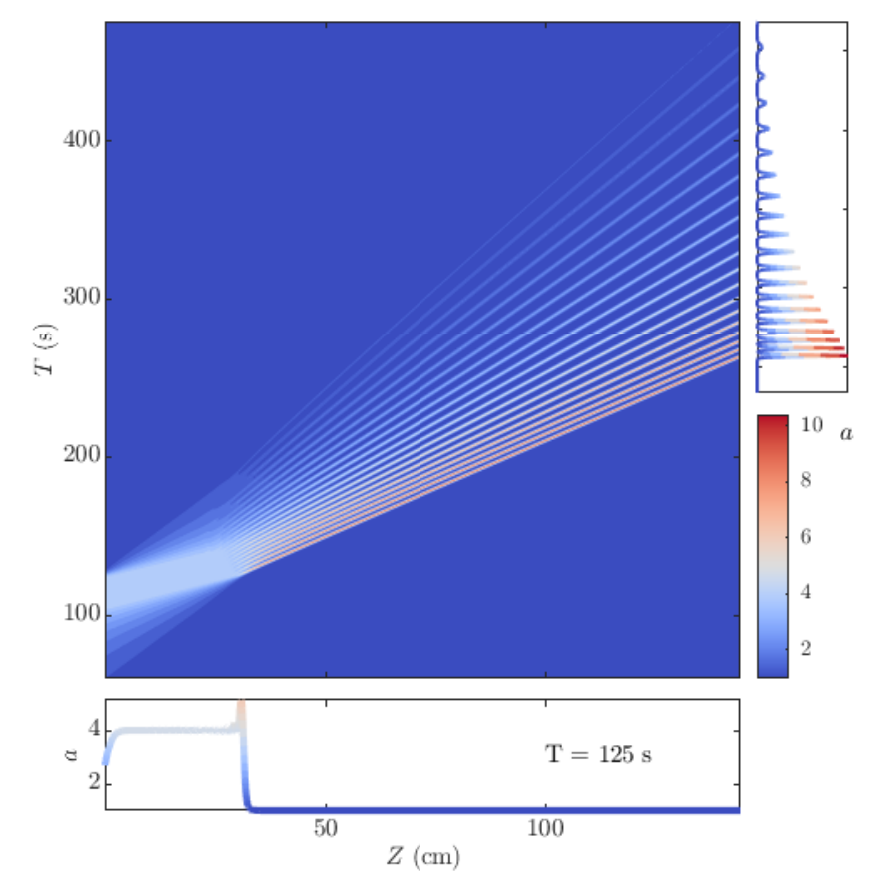}}
  \caption{Evolution of the box profile resulting in a train of
    solitary waves.  (a) Experiment.  (b) Conduit equation numerical
    simulation.  Both experiment and numerical simulation utilized the
    boundary condition \eqref{eq:box}.}
  \label{fig:box}
\end{figure}

\subsection{Results for Triangle, N, and Box Profiles}

Experiments were also performed for other wavebreaking profiles.
Experiments on boundary conditions for generating triangle waves
\eqref{eq:triangle} and N-waves \eqref{eq:N} showed good fidelity to
the expected shapes, as shown in Fig.~\ref{fig:GeneralizedCharExpt}.
Moreover, we obtain quantitative agreement between the predicted
characteristics (recall Fig.~\ref{fig:chars2}(c,d)) and the observed
curves of equi-area pre-breaking as shown by the dashed lines
overlayed on the cross-sectional area contour plot. For the figure, we
fit $z_b$ and $t_b$ to those found experimentally, then generated the
predicted characteristics (contours) based on these values.  We find
this fitting method is equivalent to fitting $\mu\ifl/\Delta$ to the
data, similar to what was done for the step profile.

As an application of our approach to generating desired wave profiles
at breaking, we show an experimental and numerical realization of
solitary wave fission \cite{trillo_experimental_2016} from the
long-time dynamics of a box profile in Figs.~\ref{fig:box}(a) and
\ref{fig:box}(b), respectively.  The boundary time-series for the box
profile, eq.~\eqref{eq:box}, results in an approximately
rectangular-shaped area profile with a specified width and height.
The long-time evolution of this profile results in a train of solitary
waves.  The numerical simulation in Fig.~\ref{fig:box}(b), generated
by the same boundary control procedure, exhibits excellent agreement
with the experiment in Fig.~\ref{fig:box}(a) for the experimental
volumetric flow rate $Q_0 = \SI{0.2}{\milli\liter\per\minute}$ if we
introduce the fitted values $\mu^{(i)}/\Delta =
\SI{724}{\square\centi\meter\per\minute}$ and $\mu^{(i)}/\mu^{(e)} =
0.05$ (measured values $\mu^{(i)}/\Delta =
\SI{1087}{\square\centi\meter\per\minute}$ and $\mu^{(i)}/\mu^{(e)} =
0.02$).  Future work aims to explore the post-breaking dynamics of
this and other dispersive hydrodynamic problems by leveraging the
boundary control method introduced here.

\section{Conclusion}\label{sec:Con}

While previous experimental and theoretical work on the conduit
equation and its corresponding viscous two-fluid system have primarily
focused on dispersive hydrodynamics \textit{post-breaking}, i.e., when
nonlinearity and dispersion are important, this work considers the
simpler case of very long waves where nonlinearity dominates the flow.
A simple long-wave hyperbolic model (the inviscid Burgers equation) is
used to theoretically control nonlinear wave propagation at the
interface of a viscous fluid conduit prior to wavebreaking.  The
pre-breaking validity of the hyperbolic model enables the precise
creation of desired wavebreaking profiles in the interior of the
dispersive hydrodynamic domain with only boundary control.
Characteristics are propagated backward in time from a desired wave
profile until they reach the boundary.  So long as the backward
characteristics do not overlap, it is possible to obtain a bounday
condition whose forward propagation approximately results in the
desired wavebreaking profile.

In order to compare this dispersionless theory with the
``dispersionfull'' conduit equation and experiment, we define
wavebreaking by an inflection point criterion for the wavefront's
slope.  This definition provides a bridge between the long-wave,
pre-breaking dynamics and the short-wave oscillations that emerge
post-breaking.  Most importantly, we obtain quantitative agreement
between theory, numerical simulation, and experiment with this
wavebreaking definition.  For a step profile, the observed breaking
heights and times are within $5\%$ and $2.5\%$, respectively, of their
expected values.  For more complex profiles---the triangle and N-wave
configurations---we obtain good characteristic control observed in
measured space-time contour plots.  One example of post-breaking
dispersive hydrodynamics is highlighted where a large number of
solitary waves emerges from a large box profile.  Numerical simulation
and experiment utilizing the same box profile boundary condition
result in striking agreement and motivate further analysis of the
soliton fission problem in the context of the viscous fluid conduit
system.  The relatively simple characteristic method proposed here
holds promise for other dispersive hydrodynamic media.

%

\appendix*
\section{Boundary Conditions for Wavebreaking Profiles}\label{appx:BCs}
All following profiles have a breaking time $t_b$ based on the
breaking height $z_b$ of $t_b = z_b/2$.
    
The boundary condition $a(0,t)$ resulting in an approximate step
profile for \eqref{eq:conduitND} (see Fig. \ref{fig:chars2}a)
\begin{equation}
  \label{eq:1}
  \begin{split}
    a(0,t) = \left\{\begin{array}{lcccccc}
        1 &:&  &   & t & \le & 0 \\
        (1-2t/z_b)^{-1} &:&  0 & < & t & < &\frac{(a_b - 1)}{2a_b} z_b    \\
        a_b             &:& &   & t & \ge & \frac{(a_b - 1)}{2a_b} z_b
      \end{array}\right. .
  \end{split}
\end{equation}
For a box, this profile is cut off at the time of breaking $t_b=z_b/2$
(see Fig. \ref{fig:chars2}b)
\begin{equation}
  \label{eq:box}
  \begin{split}
    a(0,t) = \left\{\begin{array}{lcccccc}
        1 &:&  &   & t & \le & 0 \\
        (1-2t/z_b)^{-1} &:&  0 & < & t & < &\frac{(a_b - 1)}{2a_b} z_b    \\
        a_b             &:&  \frac{(a_b - 1)}{2a_b} z_b & \le & t & < & \frac{z_b}{2} \\
        1               &:& &   & t & \ge & \frac{z_b}{2}
      \end{array}\right. .
  \end{split}
\end{equation}
For a right triangle with height $a_b$, width $w$, and hypotenuse
slope $m=\frac{a_b-1}{w}$, once the maximum desired height is reached,
we begin decreasing the flow rate in a way consistent with
Eq.\eqref{eq:IBEsolnA} (see Fig. \ref{fig:chars2}c)
\begin{equation}
  \label{eq:triangle}
  \begin{split}
    a(0,t) = \left\{\begin{array}{lcccccc}
        1 &:& & & t & < & 0 \\
        (1-2t/z_b)^{-1} &:& 0 & \le & t & < & \frac{(a_b - 1)}{2a_b} z_b   \\
        \frac{-mz_b+a_b}{1-mz_b+2mt} &:& \frac{(a_b - 1)}{2a_b} z_b & \le & t & < & \frac{w}{2}               \\
        1                                           &:& &   & t & \ge & \frac{w}{2}                            \\
      \end{array}\right.
  \end{split}
\end{equation}
The width-height restrictions on the triangle wave are based on having
the triangle fully in the conduit at the time of breaking as well as
the non-breaking condition $a_0'(z)<1/2t_b$
\begin{equation}
  1 \leq \frac{z_b}{w} \leq \frac{1}{a_b-1}
\end{equation}
    
For an N-wave with maximum height $a_{max}$, minimum height $a_{min}$,
width $w$, and slope $m = \frac{a_{max}-a_{min}}{w}$, we generate a
triangle wave whose final area dips below the mean flow rate down to
$a_{min}$ before returning to the mean flow (see
Fig. \ref{fig:chars2}d)
\begin{equation}
  \label{eq:N}
  \begin{split}
    a(0,t) = \left\{\begin{array}{lcccccc} 1 &:& & & t & < & 0
        \\
        (1-2t/z_b)^{-1} &:& 0 & \le & t & < & \frac{(a_{max} -
          1)}{2a_{max}} z_b
        \\
        \frac{-mz_b+a_{max}}{1-mz_b+2mt} &:& \frac{(a_{max} -
          1)}{2a_{max}} z_b & \le & t & < &
        \frac{w-(1-a_{min})z_b}{2a_{min}}
        \\
        \frac{z_b-w}{z_b-2t} &:& \frac{w-(1-a_{min})z_b}{2a_{min}} &
        \le & t & < & \frac{w}{2}
        \\
        1 &:& & & t & \ge & \frac{w}{2}
      \end{array}\right.
  \end{split}
\end{equation}
The N-wave width-height restrictions are also based on having the wave
fully in the conduit at the time of breaking and the non-breaking
condition $a_0'(z)<1/2t_b$
\begin{equation}
  1 \leq \frac{z_b}{w} \leq \frac{1}{a_{max}-a_{min}}
\end{equation}
Note when $w\to\infty$ and $a_{min}=a_{max}=a_b$, we arrive at
Eq.~\eqref{eq:BCDSW}.
\end{document}
